\documentstyle[mprocl]{article}

\def\be{\begin{equation}}
\def\ee{\end{equation}}
\def\bea{\begin{eqnarray}}
\def\eea{\end{eqnarray}}

\begin{document}

\title{STATISTICAL MODELING OF NUCLEAR SYSTEMATICS}

\author{J. W. CLARK}

\address{Department of Physics, Washington University, St. Louis, \\MO 63130
USA \\E-mail: jwc@wuphys.wustl.edu}

\author{E. MAVROMMATIS, S. ATHANASSOPOULOS, A. DAKOS}

\address{Physics Department, 
University of Athens, GR-15771 Athens, Greece\\E-mail: emavrom@cc.uoa.gr}  

\author{K. GERNOTH}

\address{Department of Physics, UMIST, P. O. Box 88, Manchester M60 1QD,
United Kingdom\\E-mail: k.a.gernoth@umist.ac.uk}


\maketitle\abstracts{
Statistical modeling of data sets by neural-network techniques
is offered as an alternative to traditional semiempirical
approaches to global modeling of nuclear properties.  New results
are presented to support the position that such novel techniques
can rival conventional theory in predictive power, if not
in economy of description.  Examples include the statistical
inference of atomic masses and $\beta$-decay halflives based
on the information contained in existing databases.  Neural
network modeling, as well as other statistical strategies
based on new algorithms for artificial intelligence, may prove
to be a useful asset in the further exploration of nuclear 
phenomena far from stability.}

\section{Introduction}

There is currently a strong incentive for the development of global
models of nuclear properties, driven by the production of many new 
nuclei at radioactive beam facilities and by the needs of complex 
reaction-network calculations involved in models of nucleosynthesis.  
At the same time, recent developments in statistical analysis and 
inference, especially those based on neural-network and other adaptive
techniques, present novel opportunities for global modeling
that exploit the rich collection of nuclear data currently
available. With suitable coding of input and output variables,
multilayer feedforward neural networks with pairwise couplings,
trained by backpropagation and related algorithms \cite{hertz,pim,hay},
are capable of learning from examples in the nuclear database and making 
predictions for properties of ``novel'' nuclides outside the training set 
\cite{badh}.  Neural network models have been developed for a number
of properties, including atomic masses \cite{np,pl,cpc}, neutron 
separation energies \cite{np}, ground-state spins and parities
\cite{minn,pl}, branching probabilities into different decay channels 
\cite{nn}, and halflives for $\beta^-$ decay \cite{luso}.  In terms 
of predictive accuracy, as measured on test nuclei not seen during 
training, neural-network models can compete with traditional 
phenomenological and semi-microscopic global models, although the 
number of adjustable parameters (connection weights) is generally 
much larger.  

Special attention is called to the study \cite{nn} in 
which neural networks were constructed to learn and predict whether 
a nuclide is stable in its ground state, and if not, to generate the 
branching probabilities among $\alpha$-, $\beta^-$-, electron-capture
(lumped with $\beta^+$-decay), and fission decay modes.  Global network 
models were created which show remarkable quantitative performance 
in comparison with the capabilities of standard theoretical approaches.  
In fitting and prediction of the probabilities of occurrence of 
the six modes (including stability) in nuclides belonging to the
training set and to the test set reserved for prediction, trained 
networks display errors below 5\% and 15\%, respectively, as measured 
by the average, over the relevant sets, of the {\it maximum} departure 
of the output branching probabilities from their target values.

Here we report improved results for two applications, namely to construction 
of the atomic-mass table and to determination of $\beta^-$-decay lifetimes.  
The procedures involved in neural-network statistical modeling based
on multilayer feedforward networks trained by example (``multilayer
perceptrons'') have been thoroughly 
documented in Ref.~4, which also discusses the strengths and
weaknesses of this approach.  Rather than repeat such information,
we shall simply indicate the places where essential modifications
are made.

The new experiments on learning and prediction of atomic masses 
with multilayer perceptrons are based on a new input coding scheme
and a modified backpropagation training algorithm that helps the 
system avoid local minima of the cost surface.  The results show
much better extrapability (i.e. predictive accuracy for new nuclei 
away from the stable valley) than has been found in earlier neural-network 
studies \cite{np,pl,cpc}.  Moreover, this improvement is achieved
with a greatly reduced number of connection weights.  These findings 
encourage a redoubled effort toward developing reliable mass predictors.  

Earlier work \cite{luso} on $\beta$-decay is also refined, to good
effect.  Multilayer feedforward networks trained to predict $\beta$-decay 
halflives from $Z$, $N$, and the $Q$-value of the decay are now 
performing within the range of accuracy attained by conventional 
quantum-mechanical models. 

\section{Global Models of Atomic-Mass Data}

A neural-network model will here consist of a collection of neuron-like 
units arranged in layers, with information flowing only in the forward
direction through connections between the units in successive layers.  
Input data (e.g.\ values of $Z$ and $N$) are encoded in the activities 
of the input-layer units; the activities of units in the succeeding 
layers are updated in sequence; and a coded version of the value 
computed by the network for the property in question (e.g.\ the mass 
excess) appears in the activities of the output-layer units.

The architecture of a given net is summarized in the notation
($I+H_1+H_2+\cdots+H_L+O)[P]$, where $I$, $H_i$, and $O$ are 
integers that indicate, respectively, the numbers of 
neuron-like units in the input layer, the $i$th intermediate 
(or ``hidden'') layer, and the output layer.  The total number 
of connection-weight and bias parameters is denoted by $P$.

Once the gross architecture (number of layers, number of units in each
layer) is specified, the behavior of the network -- in particular,
its response to each input pattern, each example from the database --
is entirely determined by the real-number weights of the connections 
between the units (and the biases of the units).  As the system is 
exposed to a set of training patterns, these weights (and biases) are
incrementally altered so as to minimize a {\it cost function}.  Ordinarily
the cost function is taken as the sum of the squared difference between
target and actual output activity, the sum being carried out over
all output units and over all patterns in the training ensemble.

The standard training algorithm accomplishing this goal is called
``vanilla'' backpropagation, a gradient-descent optimization routine 
which includes a ``momentum'' term to permit rapid learning without 
wild oscillations in weight space.  In our most recent mass studies, 
we institute a modification in the weight-update rule that recursively
allows earlier patterns of the current epoch (the current pass through 
the training set) to exert greater influence on the training than is 
the case for vanilla backpropagation.  Experience has shown that this 
new learning rule generally yields improved results (though not in 
all instances).

In supervised training of neural networks, there is always the question 
of when to terminate the process.  If the network is trained for too short 
a period, the training data will be poorly fitted.  On the other hand, 
if the training is continued too long, generalization to new examples 
(i.e., prediction) will suffer, since the ``overtrained''
network will be specialized to the peculiarities of the training set
that has been employed.  Thus some reasonable compromise
must be struck between the requirements of an accurate fit and good 
prediction.  In the current round of computer experiments, 
we have adopted the following stopping criterion.  A given training 
run consists of a relatively large number of epochs, specified beforehand.  
During such a run, we not only record the cost function for the patterns 
in the training set, we also monitor the cost function for a separate 
{\it validation set} of nuclei whose masses are known.  The ``trained'' 
network model resulting from a given run is taken as the one with 
connection weights yielding the smallest value of the cost function 
on the validation set, over the full course of the run.  While the 
members of the validation set are {\it not} used in the weight 
updates of the backpropagation learning rule, vanilla or modified, 
they clearly do influence the choice of model.  Therefore, accuracy 
on the validation set cannot strictly be regarded as a measure of 
predictive performance, although in practice it may nevertheless provide 
a useful (and probably faithful) indicator of this aspect of the 
model.  To obtain an unimpeachable measure of predictive performance, still 
a third set of examples is needed:  a {\it test set} that is {\it never} 
referred to during the training process. 

We are now ready to specify the data sets used in our current mass
studies.  The primary set is a database designated MN consisting of 
$1323$ ``old'' (O) experimental masses which the 1981 M\"oller-Nix model 
\cite{MN81} were designed to reproduce, together with $351$ ``new'' 
(N) experimental masses that lie mostly beyond the edges of the 1981 
data collection as viewed in the $N-Z$ plane.  This database, with this 
segmentation, has been used in previous mass-modeling exercises with 
neural nets.  As discussed in Ref.~12, the O and N data sets were 
formed to quantify the extrapolation capability (extrapability) of 
different global models of atomic masses.  In creating neural-network 
models, the old (O) masses constitute the training set, while the new 
(N) masses are employed as a validation set or -- in the earlier 
treatments -- as a test set (or prediction set).  In addition to the 
MN database, we make use of another data set composed of the mass-excess
values of 158 newer nuclides drawn from the NUBASE evaluation of nuclear and 
decay properties; generally these nuclides reside still further from the 
stable valley than those of the N subset.  This secondary data set, which 
we call NB, has been used exclusively as a test set; it is not consulted 
during the training phase.  A third set of examples, which also remains 
untouched during the training process, consists of 10 ``even newer'' 
nuclides of rare-earth elements; their masses have recently been measured 
with the ISOLTRAP mass spectrometer \cite{ISO}.  We label this last 
set ISO.
 
Three different input coding schemes are relevant to our considerations.
In the first \cite{np,pl,minn}, the input layer consists of sixteen ``on-off''
units having activity levels 0 or 1.  Eight units for $Z$ and eight
for $N$ serve to encode the proton and neutron numbers in binary and
permit the treatment of $Z$ and $N$ values up to 255, which is more
than sufficient to cover the interesting physical range of input
patterns $\nu =(Z,N)$.  This scheme facilitates learning of quantal 
properties (pairing, shell structure) that depend on the integral 
nature of $Z$ and $N$.  The second scheme utilizes analog coding of 
$Z$ and $N$ in terms of the activities of only two analog input neurons, 
which, however, are aided by two further on-off input units that 
encode the parity (even or odd) of $Z$ and $N$.  Thus the network is 
again given information about the integral character of $Z$ and $N$.
In the third scheme, the 16-unit binary-coding input array of the first
design is supplemented by two additional units encoding $Z$ and $N$ in
analog.

For all three choices of input coding, the mass computed by the network 
is represented by the activity of a single analog output unit.  Three 
different prescriptions have been used to scale the activities of the
analog units in the input and output layers to the interval [0,1];
the details, not particularly relevant here, will be presented elsewhere.
We need only remark that proper attention to the dynamical ranges of 
the $Z$, $N$, and mass-excess variables allows more precise study 
of new nuclei far from the stability line. 

Quantitative judgments of network performance in learning, validation, 
and prediction are made in terms of two different measures,
evaluated separately for the three data sets specified above. 
The chief quality measure is the root-mean-square deviation
$\sigma_{\rm rms}$ of the network-generated mass values from their
corresponding experimental target values.  This is of course
the standard quality measure adopted in global modeling of
atomic masses.  We also observe that it is just the
square root of the quantity that one seeks to minimize by
the backpropagation algorithm or one of its relatives.
The second quality measure considered here (``recalled patterns'')
is the number of patterns (nuclidic examples) for which the 
mass-excess value generated by the model deviates from experiment 
by no more that 5\%.

Another general question that arises in neural-network applications
is that of optimal architecture.  We have made no attempt at systematic 
optimization of architecture, following instead an empirical (or 
``trial-and-error'') approach.  However, in a few cases we have 
implemented a simple procedure for eliminating (``pruning'')
unimportant connections.  Connections whose omission does not 
increase the cost function beyond a small threshold amount are deleted,
and the resulting trimmed network then retrained \cite{pl,nn}. 
\begin{table}
\caption{Comparison of neural-network models of atomic
mass data with other models based on conventional nuclear theory.
Learning (fitting) and generalization (prediction) refer to the 
database MN[1323(O)-351(N)].\label{tab:table1}}
\vskip .3truecm
\begin{center}
\footnotesize
\begin{tabular}{|c|c|c|c|c|}
\hline\hline
Model&\multicolumn{2}{|c|}{Learning Mode}
&\multicolumn{2}{|c|}{Validation (v) or}\\
Neural Network&&&\multicolumn{2}{|c|}{Prediction (p) Mode}\\
$(O+H_1+H_2+\cdots+H_L)[P]$&$\sigma_{\rm rms}$&Recalled
&$\sigma_{\rm rms}$&Recalled\\
or Conventional&(MeV)&Patterns&(MeV)&Patterns\\
\hline
(16+10+10+10+1) [401]&0.393&1172/1323&3.575 (v)&246/351\\
$Z$ \& $N$ in binary&&&&\\
\hline
(18+10+10+10+1) [421]&0.331&1187&2.199 (v)& 272\\
$Z$ \& $N$ in binary and analog&&&&\\
\hline
(4+10+10+10+1)*[281]&0.491&1141&1.416 (v)&280\\
$Z$ \& $N$ in analog and parity&&&&\\
\hline
(4+10+10+10+1)** [273]&0.617&1095&1.209 (v)&284\\
$Z$ \& $N$ in analog and parity&&&&\\
\hline
$(4+10+10+10+1)^\dagger[281]$&0.453&1242&1.200 (v)&298\\
$Z$ \& $N$ in analog and parity&&&&\\
\hline
(18+10+10+10+1) [421]&0.828&---&5.981 (p)&---\\
$Z$ \& $N$ in binary,&&&&\\
A \&Z-N in analog&&&&\\
\hline
(4+40+1)[245]&1.068&---&3.036 (p)&---\\
\hline
M\"oller et al.$^{14}$ &0.673&---&0.735 (p)&---\\
\hline\hline
\end{tabular}
\end{center}
\end{table}

A substantial number of computer experiments have been carried out
for networks with various architectures and for networks with the same
architecture but different random choices of initial weights.
The first five rows of Table 1 give performance measures for some 
of the best network models emerging from these studies.  The
model identified with two asterisks is in fact a pruned (and
retrained) version of the network marked with a single asterisk.  
The model indicated with a cross gives the best neural-net 
representation of atomic-mass systematics created to date, judging by
the accuracy of its outputs for the N set of ``new'' nuclei.  The 
relevant performance measures for this data set -- here considered as 
a validation set -- appear in the last two columns of Table 1.   

The net $(4+10+10+10+1)^\dagger[281]$ appears to be distinctly superior 
to the neural-network models from earlier studies that used the O nuclei 
of the MN database as a training set.  The best models from two such 
investigations are included in Table 1.  The network listed in row 6, 
having architecture ($18+10+10+10+1)[421]$, was constructed by Gernoth 
et~al.~\cite{pl} using vanilla back-propagation, with binary encoding 
of $Z$ and $N$ together with analog encoding of the atomic mass number 
$A$ and the neutron excess $N-Z$.  The three-layer net ($4+40+1)[245]$ 
was provided by Kalman \cite{cpc}, who employed analog coding of $Z$ 
and $N$ and auxiliary parity units for these variables.  The input 
patterns were pre-processed by singular-value decomposition, and 
the network was trained with a Powell-update conjugate-gradient 
optimization algorithm.  

The gold standard of quality in global mass modeling is presently set by
the macroscopic/microscopic theoretical treatment of M\"oller, Nix, 
and collaborators \cite{mnetal}, for which rms values are entered
in the last row of Table 1.  Let us consider the rms error figures
for the ``cross'' network model, in comparison with this standard.
The result $\sigma_{\rm rms}=1.200$ MeV for the performance of the 
``cross'' model on the N set is roughly double 
the comparable M\"oller-Nix value.  Although this would still 
be regarded as quite respectable performance, we must recall 
that the N set was allowed to exert some influence on the training 
process.  On the other hand, the NB set of ``newer'' nuclides does 
qualify as a legitimate test set and indeed provides a strong test 
of the extrapability of the models we have developed. The value of 
$\sigma_{\rm rms}$ obtained for the NB set with network 
$(4+10+10+10++1)^{\dagger}$[281] is 1.462 MeV, which
is to be compared with the figure 0.697 MeV attained by
the FRDM macroscopic/microscopic model of Ref.~14.
Another strong test of the predictive quality of our models is possible
for 10 rare-earth nuclides of the ISO set, which are not contained 
in the O, N, or NB sets.  The rms error for these nuclides is found 
to be 0.963 MeV and 0.500 MeV for the ``cross'' network and for 
the FRDM model, respectively.

Based on these tests, it should be evident that the current 
generation of neural-network models of the mass table represents 
a significant advance toward extrapability levels competitive with 
those reached by the best traditional global models rooted in 
quantum theory.

\section{Global Models of $\beta^-$-Decay Halflives}

Neural-network statistical methodology is being applied to 
the systematics of nuclear decay, and in particular to the  
the important problem of predicting the halflives $T_{\frac{1}{2}}$ 
of nuclear ground states that decay 100\% by the $\beta^-$ mode.
A first effort in this direction is described in Ref.~10. 
We present here some results of a continuation of this work.

Since the relevant experimental halflives vary over 26 orders of
magnitude, the target variable in these studies is taken to be
$\ln T_{\frac{1}{2}}$.  Vanilla backpropagation, involving a
mean-square cost function, logistic activation functions, and 
a momentum term in the update rule is employed for on-line training 
of a variety of multilayer feedforward models.  The cost function is
\begin{equation}
C = {1 \over N} \sum_{\nu=1}^N  \left[ \ln { T_{1/2}^{\rm exp}(\nu)
\over T_{1/2}^{\rm model} (\nu) } \right]^2 \, ,
\end{equation}
where $\nu$ indexes the examples (input patterns) in the relevant
set of $N$ samples.  Each training run involves a pre-set number 
of epochs, and the weights that are kept after each run are taken 
as those yielding the best (smallest) value of the ``Klapdor'' 
error measure $\langle x \rangle_K$ achieved for the training set 
during the run.  Referring for example to Staudt et~al.~\cite{staudt}, 
this error measure is defined as
\begin{equation}
\langle x \rangle_K = {1 \over N} \sum_{\nu=1}^N x_\nu \,, \qquad
x_\nu \equiv \left[{T_{1/2}^{\rm exp} (\nu) \over T_{1/2}^{\rm model} 
(\nu) } \right]_>  \, ,
\end{equation}
where the sum is performed over the training or test set as appropriate
and the symbol $[a,b]_>$ stands for the ratio formed from its arguments 
$a$, $b$ in such a way that it is always larger than 1.

The raw database used in the beta-decay modeling experiments consists
of all pertinent data available in early 1995 from the Brookhaven National 
Nuclear Data Center, encompassing a total of 766 examples of single-mode
decay by the $\beta^-$ channel.  The halflives in this collection
range from $0.15 \times 10^{-2}$ sec for ${35 \atop 11}$Na to 
$0.2932 \times 10^{24}$ sec for ${143 \atop 48}$Cd.  In our most
recent modeling efforts, we have narrowed attention to the 
subset of cases in which the decay is from the ground state and
has a lifetime not longer than 10$^6$ y.  Thus we delete all
examples with longer lifetimes, as well as a few examples of
isomeric decays, arriving at a truncated data set of 692 nuclides,
of which 518 are reserved for training and 174 are used to test
predictive acuity.  This choice of database permits reasonable 
comparisons to be made with the results from traditional global 
models of $\beta^-$ halflives (see Refs.~16 and 17
and literature cited therein).

We implement binary coding of $Z$ and $N$ at the input layer, choosing
the same scheme as used in some of the mass models.  To the
operative bank of 16 on-off neurons, we append an additional analog 
input unit that encodes the $Q$-value of the decay as a floating-point
variable.  A single analog output unit generates the coded value 
of $\ln T_{\frac{1}{2}}$ that the network computes for the
input nuclide.
\begin{table}
\caption{Assessment of the ability of the selected neural-network model 
of type $(17+10+1)[191]$ to reproduce experimental values of $\beta^-$-decay 
half-lives, in comparison with traditional models of M\"oller 
et~al.$^{17}$ and Homma et~al.$^{16}$.  For good performance, the 
quantities $M^{(10)}=10^{\langle x \rangle_M}$ and $\sigma_M^{(10)}$ 
should be as small as possible.  
\label{tab:table2}}
\vskip .3truecm
\begin{center}
\footnotesize
\begin{tabular}{|c|r|c|r|c|c|r|c|r|c|}
\hline\hline
$T^{\rm exp}_{1/2}$
&
&\multicolumn{2}{|c|}{Learning}
&\multicolumn{2}{|c|}{Prediction}
&\multicolumn{2}{|c|}{M\"oller {\it et al.}}
&\multicolumn{2}{|c|}{Homma {\it et al.}}\\
(sec)&&$M^{(10)}$&$\sigma_M^{(10)}$&$M^{(10)}$&$\sigma_M^{(10)}$&$M^{(10)}$&
$\sigma_M^{(10)}$&$M^{(10)}$&$\sigma_M^{(10)}$\\
\hline
&o-o&1.15&2.27&2.05&2.31&0.59&2.91&1.75&4.96\\
$<1$&o-e&1.07&2.03&1.08&2.38&0.59&2.64&0.60&2.24\\
&e-e&1.61&1.71&1.79&2.71&3.84&3.08&1.15&2.36\\
\hline
&o-o&1.17&2.25&2.26&5.42&0.76&8.83&1.89&4.60\\
$<10$&o-e&1.04&1.91&1.19&2.44&0.78&4.81&0.92&3.84\\
&e-e&1.19&2.09&1.31&2.30&2.50&4.13&1.01&2.93\\
\hline
&o-o&1.18&2.18&1.76&5.19&2.33&49.19&3.15&10.51\\
$<100$&o-e&1.05&1.93&1.12&3.15&1.11&9.45&1.07&4.29\\
&e-e&1.19&1.97&0.98&2.67&2.61&4.75&1.13&3.58\\
\hline
&o-o&1.19&2.13&2.22&6.25&3.50&72.02&3.02&10.25\\
$<1000$&o-e&0.98&1.99&1.22&5.50&2.77&71.50&1.10&5.55\\
&e-e&1.14&1.95&0.93&4.78&6.86&58.48&1.39&6.10\\
\hline\hline
\end{tabular}
\end{center}
\end{table}
\%
Among the many network models constructed and tested, we single out for 
further attention a network with architecture $(17+10+1)[191]$, which 
demonstrated, overall, the best behavior in the predictive mode.  Table 2 
provides a detailed comparison of the performance of this model with 
state-of-the-art conventional global models recently developed by 
Homma~et~al.~\cite{prc54} and M\"oller et~al.~\cite{adndt66}  The
figures of merit that have been adopted for the models evaluated in
this table are those quoted by M\"oller et~al.~\cite{adndt66}, namely
the quantity $M^{(10)} = 10^M$ derived from the mean error
\begin{equation}
M = {1 \over N}\sum_{\nu=1}^N \log_{10} r_\nu \,,  \qquad
r_\nu \equiv { T_{1/2}^{\rm model}(\nu) 
\over T_{1/2}^{\rm exp}(\nu) } \, ,
\end{equation}
along with the quantity $\sigma_M^{(10)}=10^{\sigma_M}$ derived
from the standard deviation 
$$  
\sigma_M = \left[ {1 \over N} \sum_{\nu = 1}^N
            (r_\nu - M)^2 \right]^{1/2} 
$$
from the mean.  Ideally, both measures should be as small as possible.  
The comparison is broken down according to odd-odd, odd-even/even-odd, 
and even-even nuclear subclasses and according to different experimental 
lifetime ranges.  The level of performance displayed by the neural-network 
model is similar to (and in some cases better than) that of the traditional 
models.  Two caveats should accompany any appraisal of the relative 
merits of neural-net and conventional approaches.  On the one hand, 
comparison is hindered by the lack of a clear distinction between the 
aspects of fitting and prediction in the traditional treatments; and 
on the other, by the fact that the neural-network model has many more 
adjustable parameters than the traditional models.  At any rate, the 
good performance of the $17+10+1$ network model is clearly demonstrated 
in detailed comparisons with the experimental $\beta^-$ halflives of 
a set of 10 Cu isotopes and of a set of 10 nuclides found on or 
near the r-process path.  Typically, agreement well within an order 
of magnitude is obtained -- and usually within a factor 2.

These new findings suggest that it may be fruitful to seek
further improvements of network performance in the $\beta$-lifetime 
problem and to extend the approach to other decay modes, 
notably $\alpha$-decay.

\section*{Acknowledgments}
This research has been supported in part by the U. S. National 
Science Foundation under Grant No.\ PHY-9900173.


\end{document}